\documentclass[preprint,aps]{revtex4-1}

\usepackage{graphicx}

\begin{document}

\title{Evidence of Electron-Hole Imbalance in WTe$_2$ from High-Resolution Angle-Resolved Photoemission Spectroscopy}
\author{Chenlu Wang$^{1,5}$, Yan Zhang$^{1,5}$, Jianwei Huang$^{1,5}$, Guodong Liu$^{1,5,*}$, Aiji Liang$^{1}$, Yuxiao Zhang$^{1}$, Bing Shen$^{1,5}$, Jing Liu$^{1,5}$, Cheng Hu$^{1,5}$, Ying Ding$^{1,5}$, Defa Liu$^{1}$, Yong Hu$^{1,5}$, Shaolong He$^{1}$, Lin Zhao$^{1}$, Li Yu$^{1}$, Jin Hu$^{2}$, Jiang Wei $^{2}$, Zhiqiang Mao$^{2}$, Youguo Shi$^{1}$, Xiaowen Jia$^{3}$, Fengfeng Zhang$^{4}$, Shenjin Zhang$^{4}$, Feng Yang$^{4}$, Zhimin Wang$^{4}$, Qinjun Peng$^{4}$, Zuyan Xu$^{4}$, Chuangtian Chen$^{4}$, X. J. Zhou$^{1,5,6,*}$}
\affiliation{
\\$^{1}$Beijing National Laboratory for Condensed Matter Physics, Institute of Physics, Chinese Academy of Sciences, Beijing 100190, China.
\\$^{2}$Department of Physics and Engineering Physics, Tulane University, New Orleans, Louisiana 70118, USA
\\$^{3}$General Course Department, Military Transportation University, Tianjin 300161, China.
\\$^{4}$Technical Institute of Physics and Chemistry, Chinese Academy of Sciences, Beijing 100190, China.
\\$^{5}$School of Physical Sciences, University of Chinese Academy of Sciences, Beijing 100190, China.
\\$^{6}$Collaborative Innovation Center of Quantum Matter, Beijing 100871, China.
\\$^{*}$Corresponding author: gdliu\_arpes@iphy.ac.cn, XJZhou@iphy.ac.cn.
}
\date{May 23, 2017}

\begin{abstract}

 WTe$_2$ has attracted a great deal of attention because it exhibits extremely large and non-saturating magnetoresistance. The underlying origin of such a giant magnetoresistance is still under debate. Utilizing laser-based angle-resolved photoemission spectroscopy with high energy and momentum resolutions, we reveal the complete electronic structure of WTe$_2$. This makes it possible to determine accurately the electron and hole concentrations and their temperature dependence. We find that, with increasing the temperature, the overall electron concentration increases while the total hole concentration decreases. It indicates that the electron-hole compensation, if it exists, can only occur in a narrow temperature range, and in most of the temperature range there is an electron-hole imbalance. Our results are not consistent with the perfect electron-hole compensation picture that is commonly considered to be the cause of the unusual magnetoresistance in WTe$_2$. We identified a flat band near the Brillouin zone center that is close to the Fermi level and exhibits a pronounced temperature dependence. Such a flat band can play an important role in dictating the transport properties of WTe$_2$. Our results provide new insight on understanding the origin of the unusual magnetoresistance in WTe$_2$.
\end{abstract}


\maketitle

\newpage

Recently, WTe$_2$, a typical kind of transition metal dichalcogenides (TMDs), has attracted considerable attention and initiated intensive study, since it manifests many novel and intriguing physical properties including extremely large magnetoresistance (XMR)\cite{1},  type II Weyl Fermion\cite{22,23,24,25,26}, pressure-induced superconductivity\cite{18,19}, two-dimensional topological insulator in monolayer\cite{32}, and so on\cite{20,21,27,28,29,30}. In WTe$_2$, the magnetoresistance is reported to reach an order of at least 10$^{5}$$\%$$\sim$10$^{6}$$\%$ at low temperature and remains quadratic up to a field of 60 Tesla with no indication of saturation. However, the exact origin of the unusual magnetoresistance is still under debate\cite{12,13,14,15,16,17}. On the basis of band structure calculations and two-fluid model analysis, a perfect electron-hole compensation mechanism was proposed to account for the extremely large magnetoresistance and its unsaturated behavior. WTe$_2$ is regarded as a perfect semimetal that shows a small overlap between the valence-band and conduction-band states with an equal number of hole and electron carriers\cite{1,7,8,9,10,11}. It has become a mainstream mechanism after many magneto-transport and angle-resolved photoemission (ARPES) experiments were performed, though different groups reported rather inconsistent results even with the same kind of technique. However, so far direct experimental verification on the perfect compensation of electrons and holes in WTe$_2$ is still lacking. This requires an accurate and complete measurement on the band structure and Fermi surface of WTe$_2$.

The precise determination on the electronic structure of WTe$_2$ is challenging because of its multiple Fermi pockets that are tiny and located in a narrow momentum space, complications of bulk bands and surface states, and the three-dimensional nature of the electronic structure. Using the existing ARPES results it is hard to provide a conclusive answer to whether the electron and hole carriers are compensated for or not in WTe$_2$\cite{7,12,27}. In this paper, we have carried out high-resolution ARPES measurements on WTe$_2$ at different temperatures to examine on the origin of its extremely large magnetoresistance. Utilizing laser-based ARPES with high energy and momentum resolutions, we reveal the complete electronic structure of WTe$_2$. This makes it possible to determine accurately the electron and hole concentrations and their temperature dependence. We find that, with increasing the temperature, the overall electron concentration increases while the total hole concentration decreases. It indicates that the electron-hole compensation, if exists, can only occur in a narrow temperature range, and in most of the temperature range there is an electron-hole imbalance. Our results are not consistent with the perfect electron-hole compensation picture that is commonly considered to be the cause of the unusual magnetoresistance in WTe$_2$. We identify a flat band near the Brillouin zone center that is close to the Fermi level and exhibits a pronounced temperature dependence. Such a flat band can play an important role in dictating the transport properties of WTe$_2$. Our results provide new insight on understanding the origin of the unusual magnetoresistance in WTe$_2$.

The ARPES measurements were performed using our newly developed laser-based ARPES system equipped with a 6.994 eV vacuum-ultra-violet (VUV) laser a the time-of-flight electron energy analyzer (ARToF10K by Scienta Omicron)\cite{24}. The unique capability of our ARPES system, including simultaneous coverage of two-dimensional momentum space and high energy and momentum resolutions, made it possible to get obtain high resolution ARPES data on WTe$_2$. Figure 1 shows the temperature dependence of the measured Fermi surface, with the original data, its second derivative image and its extracted contour displayed in Fig. 1(a), 1(b) and 1(c), respectively. The corresponding temperature dependence of the energy bands along a few typical momentum cuts is shown in Fig. 2 and 3. These results are robust and highly reproducible by measuring on different samples, or by measuring the same sample during warming-cooling cycles.

According to our analysis of the measured constant energy contours and the band structures, combined in a comparison with the band structure calculations from our previous ARPES studies\cite{24}, the observed Fermi surfaces of WTe$_2$ can be summarized as follows: as shown in the left-most panel of  Fig. 1(c). (1) Four hole pockets are identified, labeled as $\alpha$, $\beta$ and nearly degenerate $\gamma$, and $\gamma$'; (2) Two nearly degenerate electron pockets can be resolved as $\epsilon$ and $\delta$; (3) A flat band around the $\Gamma$ point is resolved with its band top being about 5 meV below the Fermi level in our high resolution data (also see Fig. 3). (4) A prominent V-shaped SS1 Fermi surface segment can be easily resolved. Since the observed unusual magnetoresistance in WTe$_2$ represents a bulk property, it is usually believed that the surface state has very little contribution to the observed magnetotransport property. Therefore, we are not going to dwell on this feature in the following.

The electron-hole compensation mechanism is proposed to account for the extremely large magnetoresistance in WTe$_2$\cite{1}. Examination on the picture requires a precise determination on the area of each electron pocket and hole pocket, together with their temperature dependence. It is clear from Fig. 1 that, with increasing the temperature, the overall hole pockets shrink, while the electron pockets slightly expand. To determine the area of each Fermi pocket quantitatively, we plot the Fermi pocket contours in Fig. 1(c) by tracking the locus of the Fermi pockets shown in Fig. 1(b) that are the second derivative image with respect to momentum. Our high quality data allow us for the first time to depict the shape of all the Fermi pockets. Then we can accurately determine the area of each pocket, as shown in Fig. 4(a) and 4(b) for electron pockets and hole pockets, respectively. It is clear that the area of the two electron pockets increases with the temperature while the area of the four hole pockets decreases. We note that a Lifshitz transition was reported in WTe$_2$ at 160 K where the hole pockets disappear\cite{27}. From our present data in Fig. 1, we can still see the presence of the hole pockets at 165 K that are not consistent with such a Lifshitz transition. This difference might be due to different k$_z$ we measured and/or slightly different doping levels in the measured WTe$_2$ samples.

Figure 2 shows the temperature dependence of the band structures for WTe$_2$ measured along three typical momentum cuts. Figure 2(a) shows the bands measured at different temperatures along the cut 1, i.e., the $\Gamma$X direction (its location is shown in Fig. 2(i) as marked by the red arrow), from which the corresponding bands that contribute to the bulk electron pockets and hole pockets can all be seen. To keep track of the temperature evolution of the Fermi momenta for the observed bands, we plot the momentum distribution curves (MDCs) at the Fermi level measured at different temperatures in Fig. 2(d), with the black and red arrows pointing to k$_F$s of the hole- and electron-like bands, respectively. It is clear that the distance between the two Fermi momenta of the hole-like bands becomes smaller with increasing the temperature, while the Fermi momentum of the electron-like band slightly moves to a large value. Figure 2(b) shows the bands measured at different temperatures that only cut the electron pocket perpendicular to the $\Gamma$X direction (cut 2 in Fig. 2(i)). The corresponding MDCs at the Fermi level are shown in Fig. 2(e). With increasing the temperature, a slight increase of the Fermi surface along this momentum cut can be seen. Figure 2(c) shows the bands that only cut the hole pocket (cut 3 in Fig. 2(i)). The corresponding MDCs at the Fermi level are shown in Fig. 2(f). With increasing the temperature, an obvious shrinking of the Fermi surface along this momentum cut can be seen. The temperature evolution of the band structures shown in Fig. 2 is consistent with the temperature-dependent Fermi surface evolution in Fig. 1. They all indicate that, with increasing the temperature, the hole pockets exhibit an obvious shrinking while the electron pockets show a slight increase in WTe$_2$.

In the two-band model, the transport properties of materials are dictated not only by the concentration of charge carriers, but also by the mobility of the carriers. In ARPES measurements, the measured scattering rate can be closely related to the mobility of the carriers, and they are inversely proportional to each other. To derive the temperature dependence of the scattering rate for the Fermi pockets in WTe$_2$, we plot the photoemission spectra (energy distribution curves, EDCs) at different temperatures in Fig. 2(g) for the hole pockets and in Fig. 2(h) for the electron pockets, at two representative momentum positions marked as spots a and b in Fig. 2(i), respectively. The corresponding symmetrized EDCs are shown in the right panels in Fig. 2(g) and 2(h). The extracted EDC widths, which are related to the scattering rate, are plotted in Fig. 4(d) as a function of temperature. It is found that the hole pockets and electron pockets show comparable scattering rates, and both of them increase with increasing the temperature.

Figure 3 focuses on the temperature evolution of the bands measured near the Brillouin zone center $\Gamma$ point. Here the bands are measured along two perpendicular momentum cuts in the $\Gamma$X (cut 2) and $\Gamma$Y (cut 1) directions. For Fig. 3(b) measured along $\Gamma$Y (cut 1) direction,  the band structures at $\Gamma$ consists of two hole-like bands: one is the narrow band with its top at $\sim$5 meV below the Fermi level at 20K\cite{24}, while the other is a broad band with its top at $\sim$55 meV below the Fermi level that is composed of four nearly degenerate bands as seen in the band calculations\cite{24}. These two bands are rather anisotropic in the momentum space. When measured along the $\Gamma$X direction, they both become quite flat, as seen in Fig. 3(c). Figure 3(d) shows EDCs at the $\Gamma$ point where the two peaks in EDCs correspond to the two bands observed. With increasing the temperature, both the flat band and the high binding energy broad band shift to higher binding energy, as seen in Fig. 4(e). The top of the flat band shifts from 5 meV binding energy at 20 K to 30 meV at 165K while the broad band shows a similar shifting down from 55 meV binding energy at 20K to 72 meV at 165K. Moreover, the peak intensity of the flat band obviously decreases with increasing temperature, as seen from the EDCs at the $\Gamma$ point in Fig. 3(d). Figure 3(e) shows the momentum-integrated EDCs around the momentum region of the $\Gamma$ point covering the flat band. Here we find that the original data and the Fermi-distribution-function removed data show slight difference thus we only show the original data in Fig.3(d) and 3(e). At low temperature, the flat band is very close to the Fermi level which contributes some integrated spectrum weight at the Fermi surface mapping at $\Gamma$, as seen in Fig. 1. With increasing the temperature, the flat band shifts to higher binding energy, accompanied by the reduction of its peak intensity (Fig. 3(d)), leading to spectral weight reduction near the Fermi level in the integrated EDCs (Fig. 3(e) and 4(f)). The broad hole-like band also shows a similar spectral weight reduction during warming up.

The quantitative determination of the area for each electron pocket (Fig. 4(a)) and each hole pocket (Fig. 4(b)) offers an opportunity to check on the electron-hole balance in WTe$_2$ at this specific k$_z$. Figure 4(c) shows the total areas of the electron pockets (red circles) and hole pockets (blue circles) which correspond to carrier concentrations of electrons and holes in WTe$_2$. With increasing the temperature, the total number of holes shows a pronounced decrease while the total number of electrons exhibits a slight increase. The opposite trend of the temperature evolution of the electrons and holes causes the result that only at one temperature, here at 135 K, that the total areas of electrons and holes are the same. For all the other temperatures, they are different. In particular, the difference becomes larger with decreasing the temperature. At 20 K, the area of the hole pocket is more than twice that of the electron pocket.

We note that the electronic structure of WTe$_2$ shows a clear three-dimensional k$_z$ effect, therefore, one should take three-dimensional Fermi surface into consideration when examining the electron-hole compensation picture\cite{1,23,33}. For the Fermi surface sheets we measured at a given photon energy, they represent a momentum cut of the three-dimensional Fermi surface at a given k$_z$. One major finding of our work is that there is an overall Fermi level shift with temperature, from both the temperature dependence of the Fermi surface (Fig. 1) and band structures (Figs. 2 and 3). It is also consistent with the work reported before\cite{7,27}. In this case, with increasing the temperature, the Fermi level shows an overall upward shift, which gives rise to an increase of the electron pockets and a concomitant reduction in the hole pockets. This is true for the two-dimensional Fermi pockets at a given k$_z$, and it is also true for the three-dimensional electron pockets and hole pockets in WTe$_2$. Therefore, one may expect that the three-dimensional hole concentration and electron concentration in WTe$_2$ show opposite trends of evolution with temperature, similar to those shown in Fig. 4(c). In this case, the perfect electron-hole compensation, if exists, can only occur at one temperature, and in the rest of the temperature range, the electron and hole concentrations are different and their difference becomes larger when moving away from that particular temperature. If the particular temperature is at a finite temperature like 100 K, it will apparently violates the electron-hole compensation as the origin of the unusual magnetoresistance. Only if accidentally the electron-hole compensation temperature is at 0 K can the picture explain the observed large magnetoresistance at low temperature and its disappearance at high temperature. Although this is quite unlikely, more work needs to be carried out to find out whether there is a perfect electron-hole compensation in WTe$_2$ and at what temperature it may occur.

The chemical potential of the three dimensional Fermi surface shifts with temperature, but the shift is slight, which can be less than 1 meV with increasing the temperature from very low temperature to room temperature. Thus the reason why the Fermi level shift with temperature is that the density of state of the occupied state near the Fermi level is different from that of the unoccupied state. The Fermi level shift with temperature, which causes the opposite evolution trend of the temperature dependence for the electrons and holes, is the major cause of electron-hole imbalance in WTe$_2$. These cast doubt on the validity of the electron-hole compensation picture in understanding the extremely large magnetoresistance in WTe$_2$. By comparing the measured bands and Fermi surfaces with the calculated ones\cite{24}, spin degeneration of the observed Fermi pockets are mostly removed, due to very strong spin orbital coupling in WTe$_2$. The backscattering suppression caused by such an effect may also play an important role on magnetorersistance as suggested before\cite{12}. With increasing the temperature, the magnetoresistance effect rapidly becomes suppressed at high temperature in WTe$_2$\cite{1}. Such behavior was attributed to the lost balance between electrons and holes due to the thermal excitation of the high binding energy bands at $\sim$50 meV below the Fermi level\cite{7}. This is quite unlikely because it is usually impossible for such a deep band to substantially contribute to the electronic transport at the limited temperature. On the contrary, the near-E$_F$ flat band we have observed may play an important role in affecting the transport properties of WTe$_2$. The flat band is quite close to the Fermi level at low temperature, shifts away from the Fermi level with increasing the temperature, accompanied by a loss of its spectral weight near the Fermi level. Such a temperature-dependent evolution follows the temperature-dependent magnetoresistance in WTe$_2$, thus the effect of this flat band on the unusual magnetoresistive behavior should be taken into consideration.

In summary, by taking high-resolution ARPES measurements, for the first time, we have revealed the complete electronic structure of WTe$_2$. Our results show that, with increasing the temperature, the Fermi level shifts upwards, causing an increase in the electron concentration and a concomitant reduction of the hole concentration. This indicates that the perfect electron-hole compensation, if it exists, can only occur in a narrow temperature region. In the rest of wide temperature range there is an electron-hole imbalance in WTe$_2$. These results ask for re-examination on the perfect electron-hole compensation picture as the main cause of the extremely large magneto-resistance in WTe$_2$. We observed a flat band near the Brillouin zone center that is close to the Fermi level at low temperature, and get suppressed at high temperature. This flat band may play an important role in dictating the transport properties of WTe$_2$. Our results provide important information in understanding the unusual magnetoresistance in WTe$_2$ that calls for further efforts to clarify its exact origin.

\vspace{3mm}

\noindent {\bf Acknowledgement}\\

Supported by the National Natural Science Foundation of China under Grant No 11574367, the National Basic Research Program of China under Grant Nos 2013CB921904 and 2015CB921300, the National Key Research and Development Program of China under Grant No 2016YFA0300600, the Strategic Priority Research Program (B) of the Chinese Academy of Sciences under Grant No XDB07020300, and the US Department of Energy under Grant No DE-SC0014208.

\vspace{3mm}

\renewcommand\figurename{Fig.}

\newpage

\begin{figure*}[tbp]
\begin{center}
\includegraphics[width=1.0\columnwidth,angle=0]{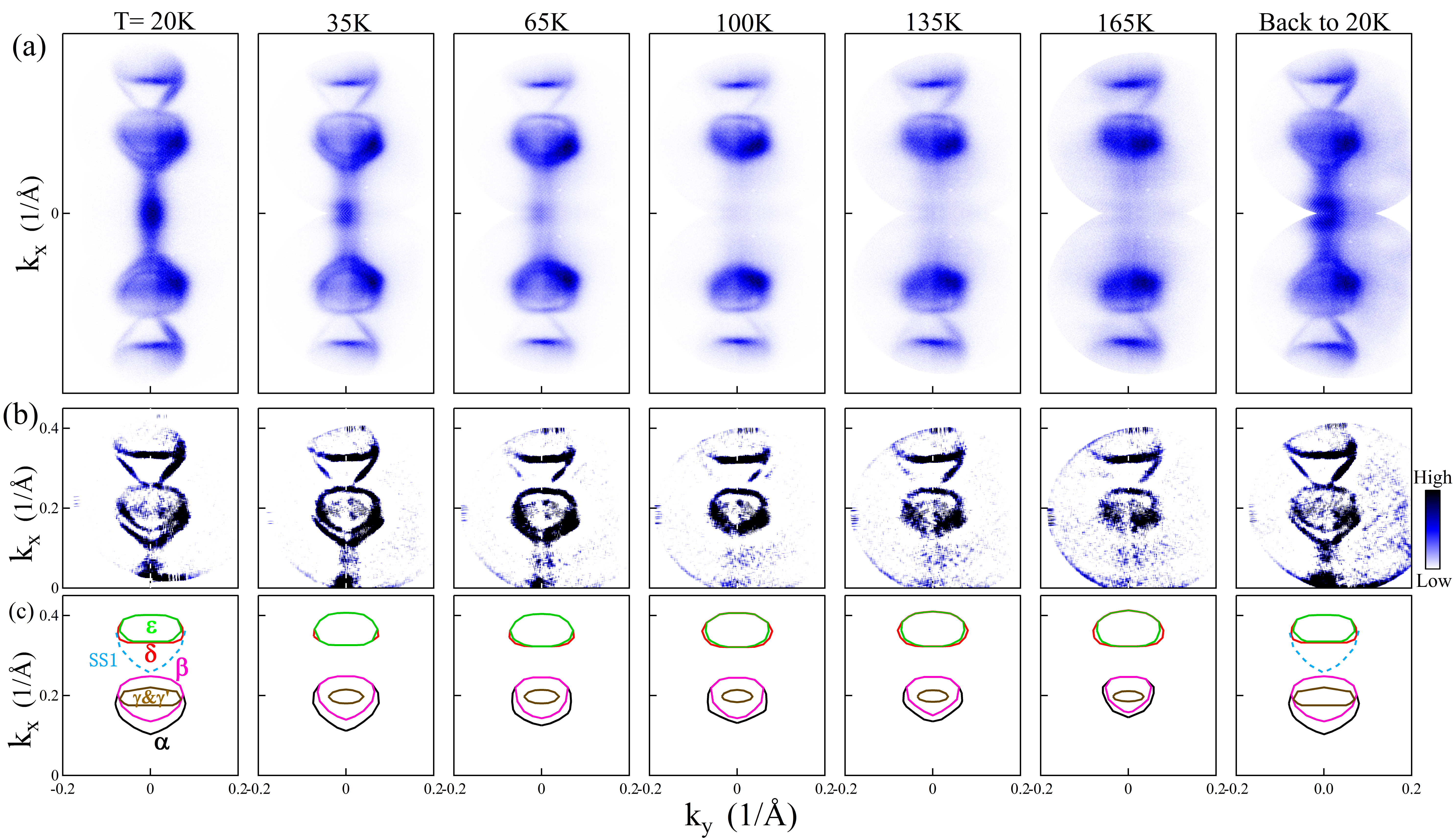}
\end{center}
\caption{Fermi surface of WTe$_2$ measured at different temperatures. (a) Fermi surface evolution with temperature when WTe$_2$ is warmed up from 20 K to 165 K and back to 20K. The Fermi surface mapping is obtained by integrating the spectral weight within an energy window of [-5 meV, +5 meV] with respect to the Fermi level. (b) The second derivative of the original images corresponding to (a) with respect to the momentum. (c)The extracted Fermi surface sheets corresponding (a) and (b). The solid lines represent the bulk Fermi pockets while the dashed lines represent the surface state segments. Different pockets are drawn with different colors.
}
\end{figure*}

\begin{figure*}[tbp]
\begin{center}
\includegraphics[width=1.0\columnwidth,angle=0]{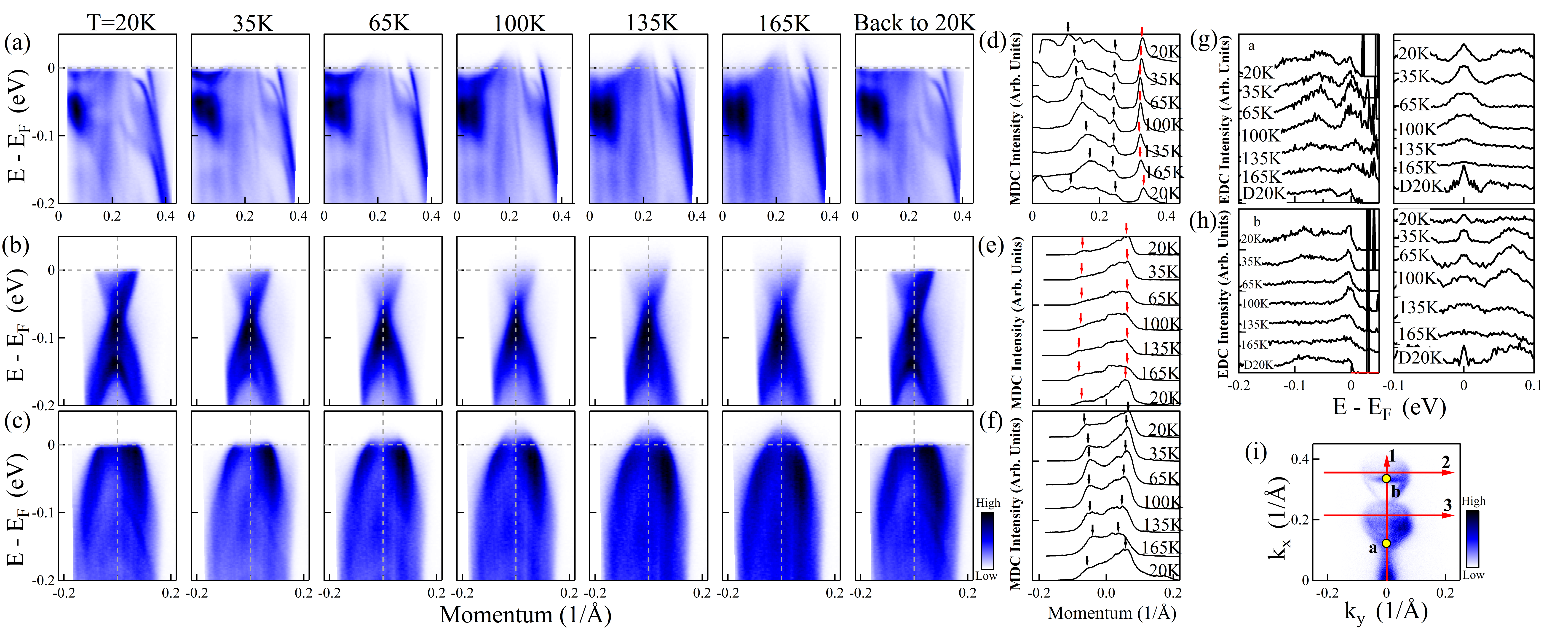}
\end{center}
\caption{Temperature evolution of the band structures in WTe$_2$ measured along typical momentum cuts. (a)-(c) Band structures measured along cuts 1, 2 and 3 at different temperatures. The locations of the momentum cuts are marked by red lines in (i). (d)-(f) The momentum distribution curves (MDCs) at the Fermi level measured at different temperatures in (a)-(c). Black arrows mark the hole pockets and red arrows mark the electron pockets. [(g),(h)] The Energy distribution curves (EDCs) at different temperatures measured at the momenta (a) and (b), as marked  by yellow circles in (i). The left panels show the original EDCs, and the right panels show the symmetrical EDCs. (i) Fermi surface of WTe$_2$ measured at 20 K. The location of the momentum cuts are marked. In (d)¨C(h), the bottom curves correspond to the temperature back to 20 K.
}

\end{figure*}

\begin{figure*}[tbp]
\begin{center}
\includegraphics[width=1.0\columnwidth,angle=0]{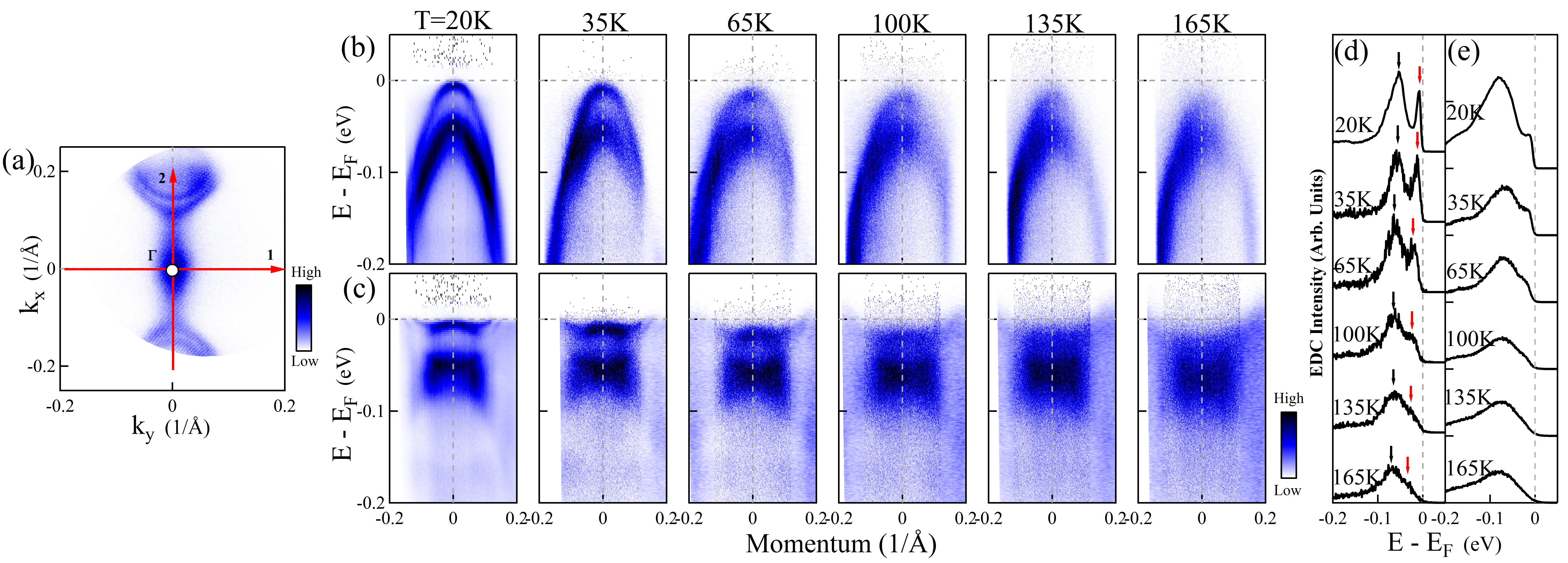}
\end{center}
\caption{Temperature evolution of the band structures in WTe$_2$ measured along $\Gamma$X and $\Gamma$Y directions. (a) Fermi surface of WTe$_2$ measured at 20 K. Two momentum cuts 1 and 2 are marked by red lines. [(b),(c)] Band structures measured at different temperatures along cuts 1 and 2, respectively. (d) The EDCs at $\Gamma$ point measured at different temperatures. The location of the EDCs are marked by white circles in (a). Red arrows mark the peaks of the flat band and black arrows mark the other band. (e) The integral EDCs of the $\Gamma$ area.
}

\end{figure*}

\begin{figure*}[tbp]
\begin{center}
\includegraphics[width=1.0\columnwidth,angle=0]{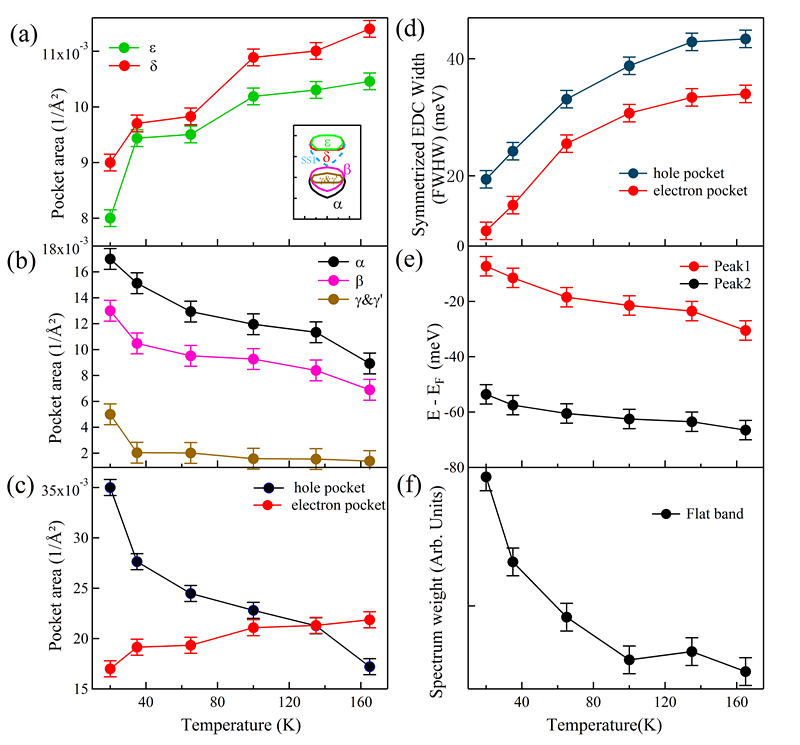}
\end{center}
\caption{Temperature dependence of quantitative parameters of electronic structure in WTe$_2$. (a) The electron pocket area at different temperatures. (b) The hole pocket area at different temperatures. (c) The total area of electron pockets and hole pockets at different temperatures. (d) The symmetrized EDC line width (full width at half maximum, FWHM) on the hole and electron pockets as marked at a and b points in Fig. 2(i). (e) The peak position of the flat band and the other band at different temperatures obtained from EDCs in Fig. 3(d). (f) The spectral weight of the flat band at different temperatures. The spectral weight is the area below the integral EDC in Fig. 3(e) within an energy window of [-30 meV, 0 meV] with respect to the Fermi level.
}

\end{figure*}

\end{document}